\documentclass[preprint]{aastex}

\usepackage{bm}
\usepackage{lineno}
\usepackage[colorlinks=true, linkcolor=blue]{hyperref}

\begin{document}
\title{New results on the onset of a coronal mass ejection from 5303{\AA} emission line observations with VELC/ADITYA-L1}

\author{R. Ramesh\altaffilmark{1}, V. Muthu Priyal\altaffilmark{1}, 
Jagdev Singh\altaffilmark{1},  
K. Sasikumar Raja\altaffilmark{1}, P. Savarimuthu\altaffilmark{1}, and Priya Gavshinde\altaffilmark{1}}

\altaffiltext{1}{Indian Institute of Astrophysics, Koramangala 2nd Block, Bangalore 560034, Karnataka, India}

\begin{abstract}
We report on the onset of a coronal mass ejection (CME) using  spectroscopic observations in 5303{\AA} coronal emission line with the Visible Emission Line Coronagraph (VELC) onboard ADITYA-L1, the recently launched first Indian space solar mission. The CME was observed on 16 July 2024 in association with a X1.9 class soft X-ray flare from heliographic location S05W85. The VELC observations were 
near the west limb of Sun during the CME. The results obtained helped to constrain the onset time of the CME. In addition, they indicate ${\approx}$50
${\approx}10$ km/s. The non-thermal velocity associated with the line broadening is ${\approx}24.87$ km/s.
\end{abstract}

\keywords{Sun: activity; Sun: corona; Sun: coronal mass ejections: CMEs; Sun: radio radiation}

\section{Introduction} \label{sec:intro}

The green line (5303{\AA}) is the brightest of all solar coronal emission
lines in the visible spectral range. It is a forbidden line in the coronal spectrum, and caused by transition between the ground state fine structure levels of Fe XIV. 
The emission peaks at a temperature of ${\approx}1.8{\times}10^{6}$\,K.
The 5303{\AA} line is an useful tracer of solar activity and loops
in the corona.
Observations of the solar corona in 5303{\AA} emission line with ground based coronagraphs are known for several 
decades \citep{Evans1957,Orrall1961,Bruzek1970,Demastus1973,Tsubaki1975,Sykora1992,
Guhathakurta1993,Ichimoto1995b,Sakurai1998,Bagala2001,
Singh1999,Singh2004,Singh2011}. The Upgraded Coronal Multi-channel Polarimeter (UCoMP; \citealp{Landi2016}), commissioned recently at the Mauna Loa Solar Observatory, is capable of observing several emission lines in the range 5303{\AA}\,-\,10830{\AA}. 
Observations in the 5303{\AA} line from space with the Large Angle Spectrometric Coronagraph C1 (LASCO-C1, \citealp{Brueckner1995}) onboard Solar and Heliospheric Observatory (SoHO) have been reported by \cite{Schwenn1997,Inhester1997,Mierla2008}. There are several observations in the green line during the total solar eclipses also \citep{Koutchmy1983,Ichimoto1995a,Rusin1995,Pasachoff2002,Pasachoff2009,
Minarovjech2003,
Singh2011,Voulgaris2012,Caspi2020,Muro2023}. But reports of dynamic events, particularly related to CMEs, are rare \citep{Dere1997,Plunkett1997,Hori2005,Suzuki2006}. 
Such observations are important to understand the thermodynamic changes in the near-Sun corona due to CMEs and other solar eruptions as shown by \cite{Boe2020} using total solar eclipse data obtained 
from multiple sites. \cite{Alzate2017} showed the presence of atypical large scale structures in coronal images obtained from two successive eclipse observations in two successive years. In both cases, the shapes of the structures were similar to that of the CME shock fronts that propagated through the same region of the corona few hours prior to the eclipse observations. In one case, the atypical structure was dominated by cool chromospheric material.
These limited studies are primarily due to the unfortunate loss of SoHO/LASCO-C1 in 1998, and practical difficulties in the observations of solar corona with ground based coronagraphs. 
In this Letter, we present comprehensive measurements of the intensity, width, and Doppler shift of the 5303{\AA} emission line before, during, and after a CME using spectroscopic observations with VELC/ADITYA-L1.

\section{The instrument} \label{sec:instr}

\begin{figure}[ht!]
\centerline{\includegraphics[width=14cm]{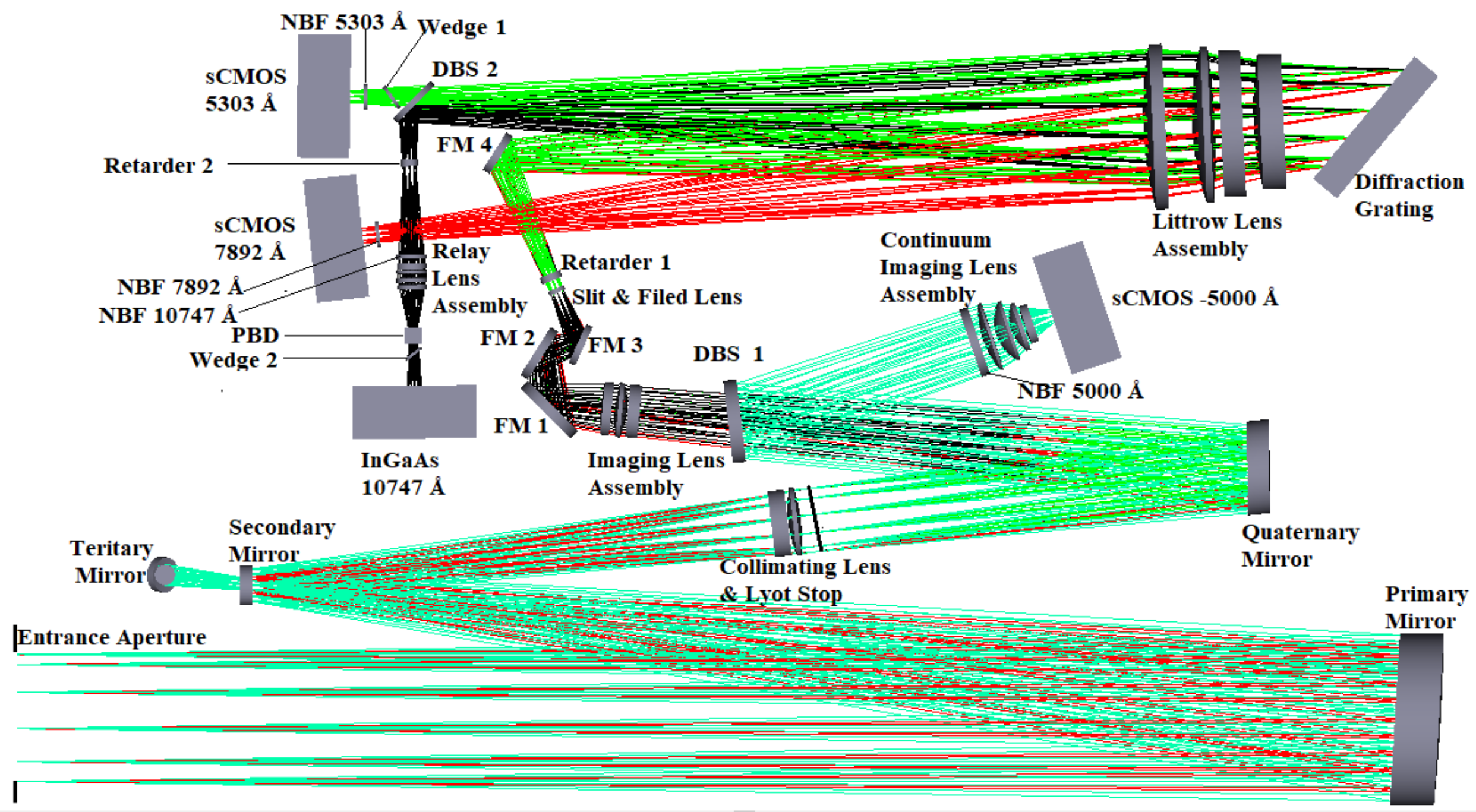}}
\caption{Optical layout of VELC}
\label{fig:figure1} 
\end{figure}

The VELC payload onboard ADITYA-L1 is an internally occulted solar coronagraph capable of carrying out simultaneous imaging (5000{\AA}) and spectroscopy (5303{\AA}, 7892{\AA}, \& 10747{\AA}) observations from close to the limb 
$r$\,=\,$\rm 1.05\,R_{\odot}$. 
The 10747{\AA} channel can be operated in either spectroscopy or spectropolarimetry mode. The field-of-view (FoV) for the imaging channel (continuum) is 1.05\,$\rm R_{\odot}$\,-\,3.0\,$\rm R_{\odot}$. 
The FoV for the spectral channels is 
1.05\,$\rm R_{\odot}$\,-\,1.5\,$\rm R_{\odot}$.  
The spectroscopic observations are carried out using four slits simultaneously. 
The continuum and spectroscopic channels have independent narrow-band filters. This helps to avoid overlap of spectra from the different slits in the case of spectral observations.
Figure \ref{fig:figure1} shows the optical layout of VELC \citep{Prasad2017,Singh2019}. Sunlight enters the payload via the entrance aperture. The primary mirror forms the image of the solar disk and corona on the secondary mirror.
Solar disk light and coronal light till $r$\,=\,1.05\,$\rm R_{\odot}$ pass through the central hole of secondary mirror and are reﬂected out of the payload by the tertiary mirror. Coronal light in the 
range $r$\,=\,1.05\,$\rm R_{\odot}$\,-\,3.0\,$\rm R_{\odot}$ is reflected by the annular region of the secondary mirror towards the quaternary mirror which in turn reflects the light towards the dichroic beamsplitter 1 (DBS1).
The latter reflects the light at wavelengths $<$5100{\AA} towards the continuum imaging lens assembly and narrow band filter (NBF) to form image of the corona in continuum radiation at the CMOS detector. The transmitted light by DBS1 at wavelengths $>$5100{\AA} passes through the imaging lens assembly. Then it is reflected by the mirrors FM1 \& FM2 mounted on a linear scan mechanism (LSM), FM3, and forms image of the corona on the four slits of the grating spectrograph. DBS2 separates the light at wavelengths 10747{\AA} and 5303{\AA}. 

\begin{figure}[ht!]
\centerline{\includegraphics[width=12.5cm]{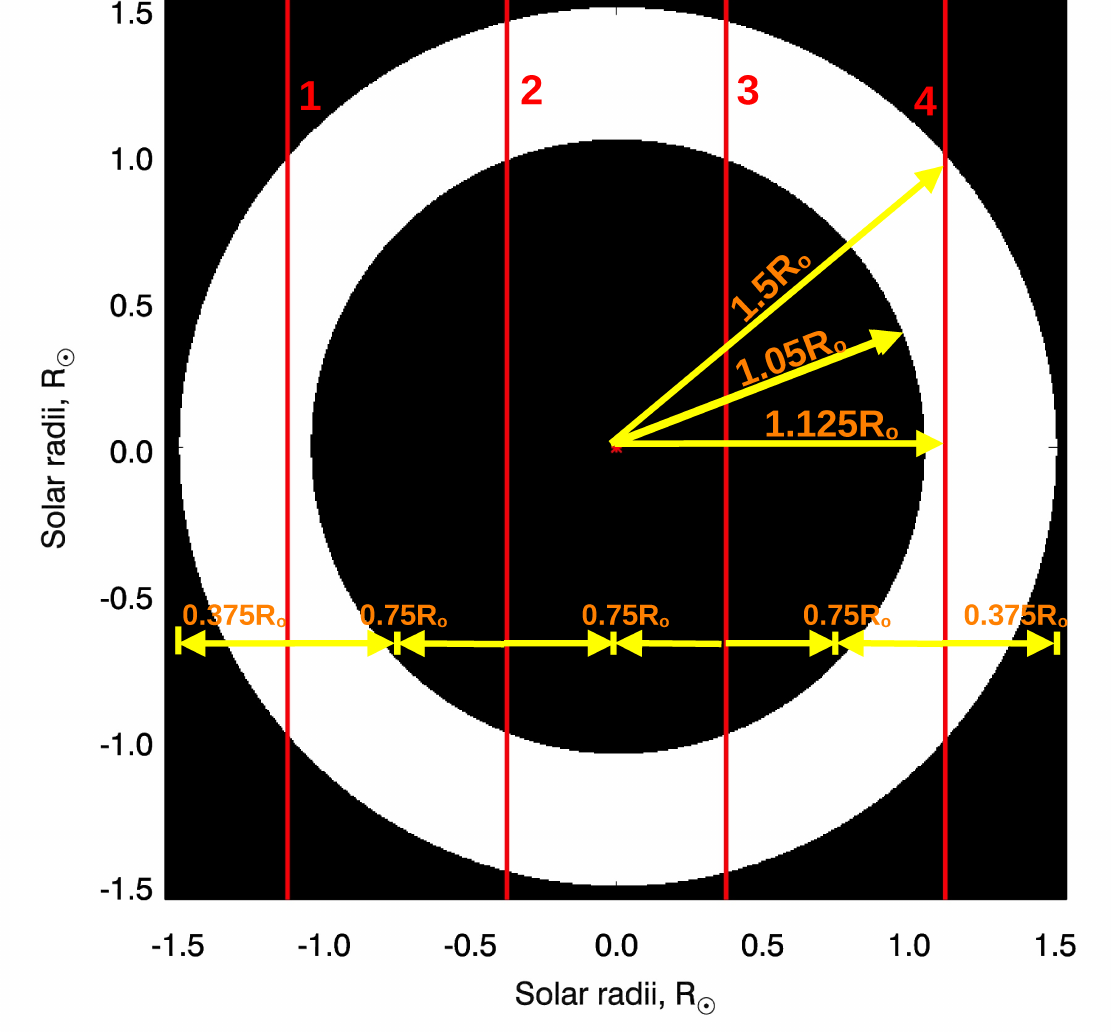}}
\caption{Schematic of the `home' positions of the four slits in the VELC spectroscopy channel. The inner filled black circle corresponds to the occulter (radius\,=\,$\rm 1.05\,R_{\odot}$). The white circular patch indicates the FoV, i.e. 1.05\,$\rm R_{\odot}$\,-\,1.5\,$\rm R_{\odot}$. Solar north and east directions are straight up and to the left, respectively.}
\label{fig:figure2} 
\end{figure}

Figure \ref{fig:figure2} shows the four slits in the spectral channel of VELC, and their FoV at the
`home' position of LSM. 
The length of each slit is along the north-south direction of Sun, and width (dispersion) is along the east-west direction of Sun. 
Some of the parameters related to the spectral channels are listed in Table \ref{tab:table1}.
The observations can be carried out in either sit \& stare mode or raster scan mode.
In the case of sit \& stare mode, fixed locations in the corona 
are observed by the four slits as shown in 
Figure \ref{fig:figure2}. Four spectra can be obtained at the same time. 
In raster scan mode the LSM is moved in steps at chosen time interval to obtain 
image of the corona by combining data obtained with the four slits. 
A distance range of 
$\rm {\pm}0.375\,R_{\odot}$ (w.r.t the above mentioned `home' position) in the east-west direction of the coronal image can be observed using each slit by moving the LSM.
All position angles
(${\theta}$, measured counter clockwise from north through east)
in the range $0^{\circ}$ to $360^{\circ}$ will be covered by the four slits together. 
The combination of Littrow lens, grating and narrow band filters help to record the spectra in three emission lines (5303{\AA}, 7892{\AA}, \& 10747{\AA}) simultaneously using three different detectors.  The spectra recorded are analyzed to derive the parameters of the emission line such as intensity, width, and Doppler velocity at the respective locations of the corona observed. The temporal variations of the above mentioned parameters  at a particular spatial location can be studied using sit \& stare observations. Their spatial variation either over a larger area in the corona or the entire corona within the FoV can be studied using raster scan observations. More details of the VELC payload can be found in \cite{Prasad2017,Prasad2023,Venkata2017,Venkata2021,Rajkumar2018,
Singh2011a,Singh2019,Singh2024,Mishra2024,Muthupriyal2024}.  

\begin{table}[!t]
\centering
\caption{Instrument parameters related to VELC spectral observations}
\label{tab:table1}
\begin{tabular}{ll}     
\noalign{\smallskip}\hline\noalign{\smallskip}
Parameter & Specifications \\
\hline
Detector pixel size & 6.5\,$\rm {\mu}m$ \\
Spatial scale & $1.25^{\prime\prime}$ per pixel \\
Slit width & 50\,$\rm {\mu}m$ (equivalent to $9.6^{\prime\prime}$) \\
Slit length & 15\,mm (equivalent to $48^{\prime}$) \\
FoV (total) & 1.05\,$\rm R_{\odot}$\,-\,1.5\,$\rm R_{\odot}$ \\
FoV (each slit position) & $\rm {\pm}1.5\,R_{\odot}\,{\times}\,0.01\,\rm R_{\odot}$ \\ 
& (spatial direction ${\times}$ dispersion direction)\\
Separation between slits & 3.75\,mm \\ 
\hline
\end{tabular}
\end{table}

\section{Observations and Data Analysis} \label{sec:obs}

\begin{figure}[ht!]
\centerline{\includegraphics[width=14cm]{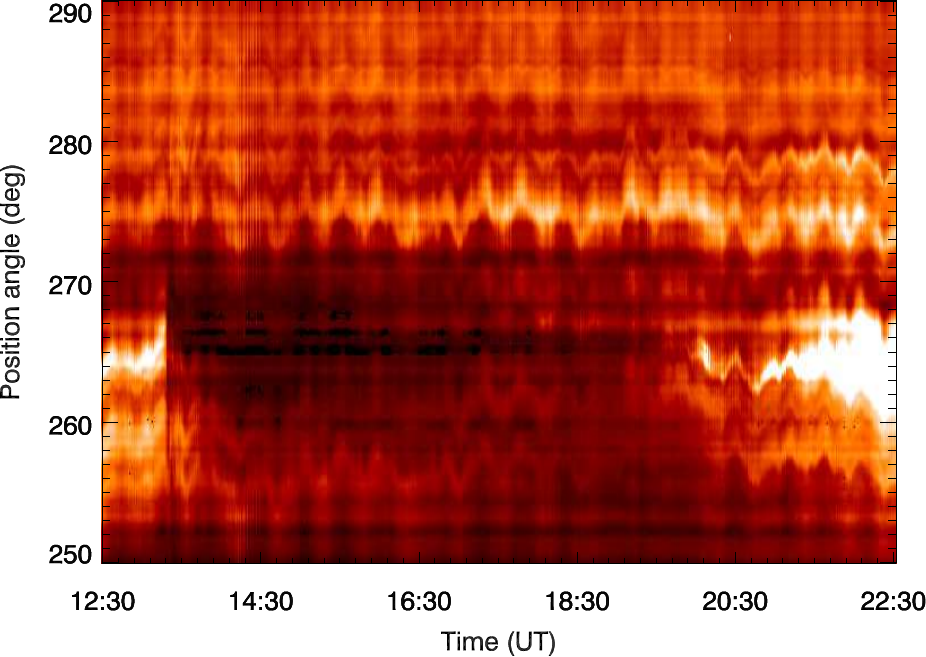}}
\caption{Temporal variations in the peak intensities of the 5303{\AA} emission line from the solar corona in the position angle range 
${\approx}250^{\circ}$\,-\,$290^{\circ}$
observed on 16 July 2024 with VELC/ADITYA-L1. 
The center of spectrograph slit 4 was positioned close to the west limb of Sun at $r$\,$\rm {\approx}1.05\,R_{\odot}$.
A sudden dimming of the intensity can be clearly observed in the position angle range
${\approx}\,260^{\circ}$\,-\,$270^{\circ}$ during the interval 
${\approx}$\,13:18\,UT\,-\,19:00\,UT. The possible reasons for the oscillatory pattern in the image are being investigated.}
\label{fig:figure3}
\end{figure}

The observations reported were carried out with VELC/ADITYA-L1 in its 5303{\AA} channel. Due to various technical issues, useful data was available in this channel only. The observational parameters related to the present work work are listed in 
Table \ref{tab:table2}. For the sit \& stare mode,
the LSM was positioned such that the center of slit 4 observes the corona at $r$\,=\,1.05\,${\rm R_{\odot}}$ in the west direction of Sun. 
Totally 703 spectra were recorded. Each spectrum was corrected for dark current, curvature of the spectra,
flat-field, and background.
After the corrections, the 
emission line profile for each spatial location along the slit {\bf was} generated. Then the emission line profile was fitted with a Gaussian to compute the peak intensity, width of the emission ine, and Doppler velocity at each spatial location along the slit. The average central position of the emission line for the background corona was used as reference to compute the Doppler velocity. All the 703 spectra were analyzed similarily to derive the above mentioned emission line parameters as a function of time, and spatial locations along the slit. The computed emission parameters for each spectrum along the  
spatial direction in the FoV were stacked as a function of time.
The image shown in Figure \ref{fig:figure3} 
was constructed using the peak intensity of the emission line. X-axis in the image corresponds to Universal Time (UT). The y-axis is the spatial extent of slit 4 corresponding to the position angle range
${\approx}250^{\circ}$\,-\,$290^{\circ}$. 
Note that the heliocentric distances of the coronal regions at position angles $250^{\circ}$ \& $290^{\circ}$ are $r{\approx}1.12\,\rm R_{\odot}$ as compared to $r=1.05\,\rm R_{\odot}$ for the center of slit 4. This is due to the straight nature of the slit.

\begin{table}[!t]
\centering
\caption{Parameters of the 5303{\AA} emission line observations on 16 July 2024}
\label{tab:table2}
\begin{tabular}{ll}     
\noalign{\smallskip}\hline\noalign{\smallskip}
Parameter & Specifications \\
\hline
Spectral dispersion & 28\,m{\AA} per pixel \\
FWHM of the narrowband filter & 6.5{\AA} \\ 
{\bf Sit \& stare mode} & \\
Location of center of slit 4 w.r.t to image & $r$\,=\,1.05$\rm R_{\odot}$ at west limb \\
Spatial binning  & 2 pixels \\
Spectral binning & 1 pixel \\
Exposure time & 5\,sec for each spectrum \\
Cadence & 51\,sec \\
Observation duration & 10\,hrs (12:30\,UT\,-\,22:30\,UT) \\
{\bf Raster scan mode} & \\
FoV & 1.05\,$\rm R_{\odot}$\,-\,1.5\,$\rm R_{\odot}$ (for all directions) \\
Step size of LSM & 
20\,${\mu}$m (image movement on slits = 40\,${\mu}$m) \\
Number of steps & 95 \\
Exposure time & 5\,sec \\
Number of spectra binned onboard  & 6 
(effective exposure time = 30\,sec)\\
Time interval between successive positions of LSM & 30\,sec \\
Duration of one raster scan & ${\approx}$47\,min \\
\hline
\end{tabular}
\end{table}

\begin{figure}[ht!]
\centerline{\includegraphics[width=14cm]{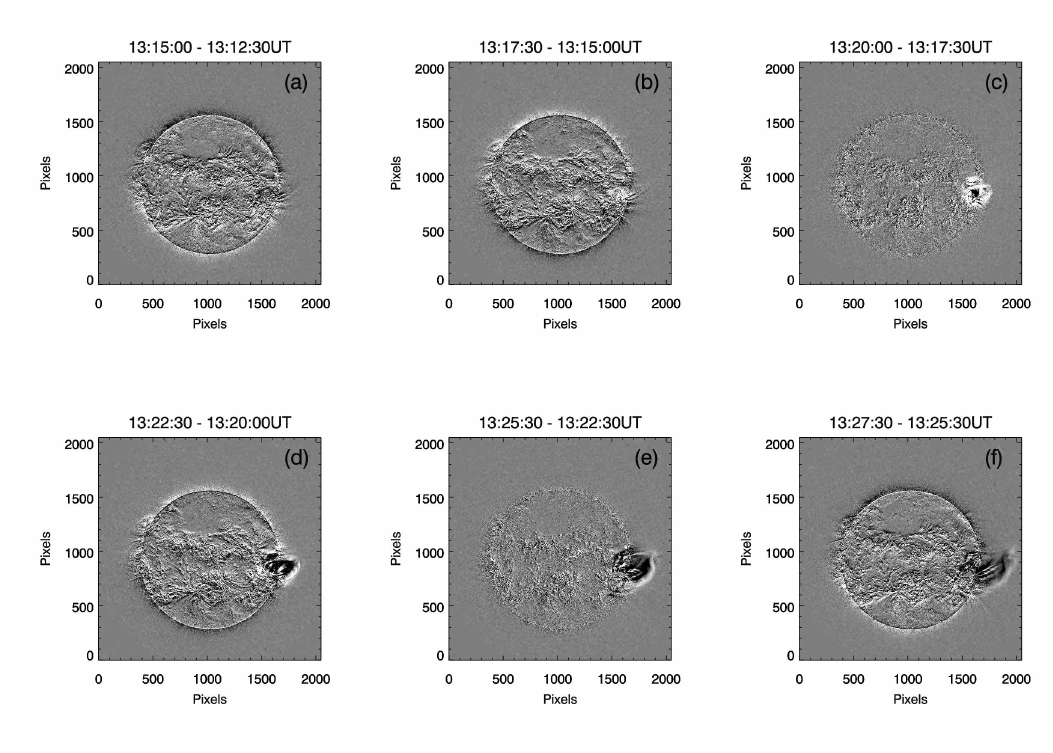}}
\caption{STEREO-A/EUVI 195{\AA} running difference images for the flare event of 16 July 2024 mentioned in the text. The observing times of the images used are mentioned above each panel.}
\label{fig:figure4}
\end{figure}

There was a sudden reduction in the intensity (coronal dimming) at 
${\approx}$13:18\,UT around position angle  
${\approx}\,265^{\circ}$.
The dimming lasted for 
${\approx}$6\,h. It was closely associated with a 1B 
H${\alpha}$ flare from AR13738 located at heliographic coordinates S06W85 in the interval 
${\approx}$13:18\,UT\,-\,14:04\,UT with maximum at 13:20\,UT. 
The position angle of the above active region is 
${\approx}264^{\circ}$.
This is in good agreement with the position angle of the dimming mentioned above.
Note that the exact onset time of the H$\alpha$ flare is uncertain due to limited available data\footnote{https://www.solarmonitor.org/data/2024/07/16/meta/noaa{\_}events{\_}raw{\_}20240716.txt}. 
Therefore, the start time of the flare could have been before 13:18\,UT. 
Figure \ref{fig:figure4} shows running difference images of the solar corona generated using observations in 195{\AA} with Extreme-UltraViolet Imager (EUVI) of the Sun-Earth Connection Coronal and Heliospheric Investigation \citep{Howard2008} onboard the Solar
Terrestrial Relationship Observatory-A (STEREO-A) around the same time. STEREO-A was at
$\rm {\approx}E19^{\circ}$ during the above epoch\footnote{https://stereo-ssc.nascom.nasa.gov/cgi-bin/make{\_}where{\_}gif}. Therefore, the above mentioned location of AR13738 corresponds to ${\approx}24^{\circ}$ inside the limb in the Earthward direction for STEREO-A view. The flare could be noticed as a brightening near pixel coordinates (1600,800) in  
the panel (b) image of Figure \ref{fig:figure4}. This panel indicates that the flare onset was during the interval 
13:15:00\,UT\,-\,13:17:30\,UT. 
The subsequent running difference images indicate an equatorward deflection (in the plane-of-sky) of the ejected coronal plasma. 
There was also a X1.9 soft X-ray flare during the interval 
${\approx}$13:11\,UT\,-\,13:36\,UT with maximum at 13:26\,UT in the Geostationary Operational Environmental Satellite (GOES) observations. 
AR13738 was very active on 16 July 2024. There were C5.0 and C4.8 flares at ${\approx}$12:57\,UT and ${\approx}$17:42\,UT, respectively. The comparitively enhanced brightness near the position angle 
${\approx}\,265^{\circ}$ in Figure \ref{fig:figure3}, before and after the dimming event, is likely because of the above mentioned flares.
Note that intensity increase due to density enhancements in the solar atmosphere appear as horizontal bright bands in the 5303{\AA} emission line intensity spectra (see e.g. \citealp{Hori2005}).

A type II radio burst was reported in the frequency range
${\approx}$\,180\,MHz\,-\,25\,MHz on 16 July 2024. Its onset time at 180\,MHz was 
${\approx}$13:21\,UT. SoHO/LASCO-C2 observed a CME centered at position angle ${\approx}273^{\circ}$ with angular width 
${\approx}92^{\circ}$ close to the flare location mentioned in the previous paragraph\footnote{http://spaceweather.gmu.edu/seeds/ql{\_}lasco/2024/07/20240716.142407.w092.v0393.p273.txt}. 
The position angle of the CME is in good agreement with that of the dimming in Figure \ref{fig:figure3}. 
The first height-time (h-t) measurement of the CME was at 
${\approx}$14:24\,UT when its leading edge was at 
$r$\,$\rm {\approx}\,4.25\,R_{\odot}$. The 2nd order (quadratic) fit to the h-t measurements of the CME in the SoHO/LASCO FoV indicate that its onset time at $r$\,=\,1\,$\rm R_{\odot}$ is ${\approx}$13:48\,UT. The corresponding time based on linear fit is ${\approx}$12:36\,UT. The estimated linear speed of the CME in the plane-of-sky is ${\approx}$393\,km/s. Its acceleration in the SoHO-LASCO FoV is  
${\approx}$\,-80.4\,$\rm m/s^{2}$. 

\begin{figure}[ht!]
\centerline{\includegraphics[width=14cm]{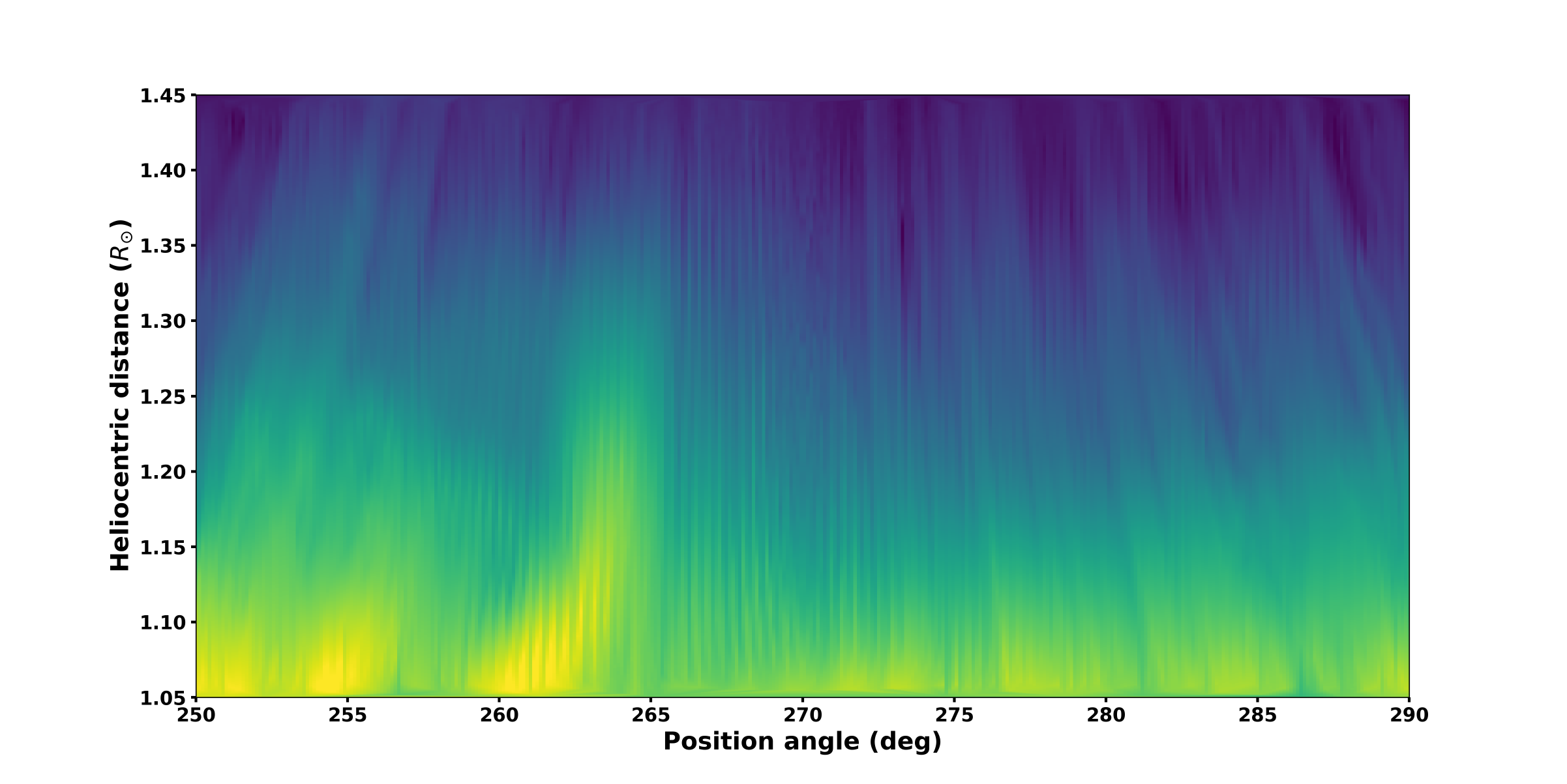}}
\caption{(r,${\theta}$) image of the solar corona at west limb of Sun constructed from the raster scan observations in 5303{\AA} emission line on 16 July 2024  during the interval ${\approx}$00:30\,UT\,-\,01:17\,UT, with VELC/ADITYA-L1. The fine lines seen in the image are due to remnant fixed pattern noise due to the detector, even after the flat field correction.}
\label{fig:figure5} 
\end{figure}

\begin{figure}[ht!]
\centerline{\includegraphics[width=14cm]{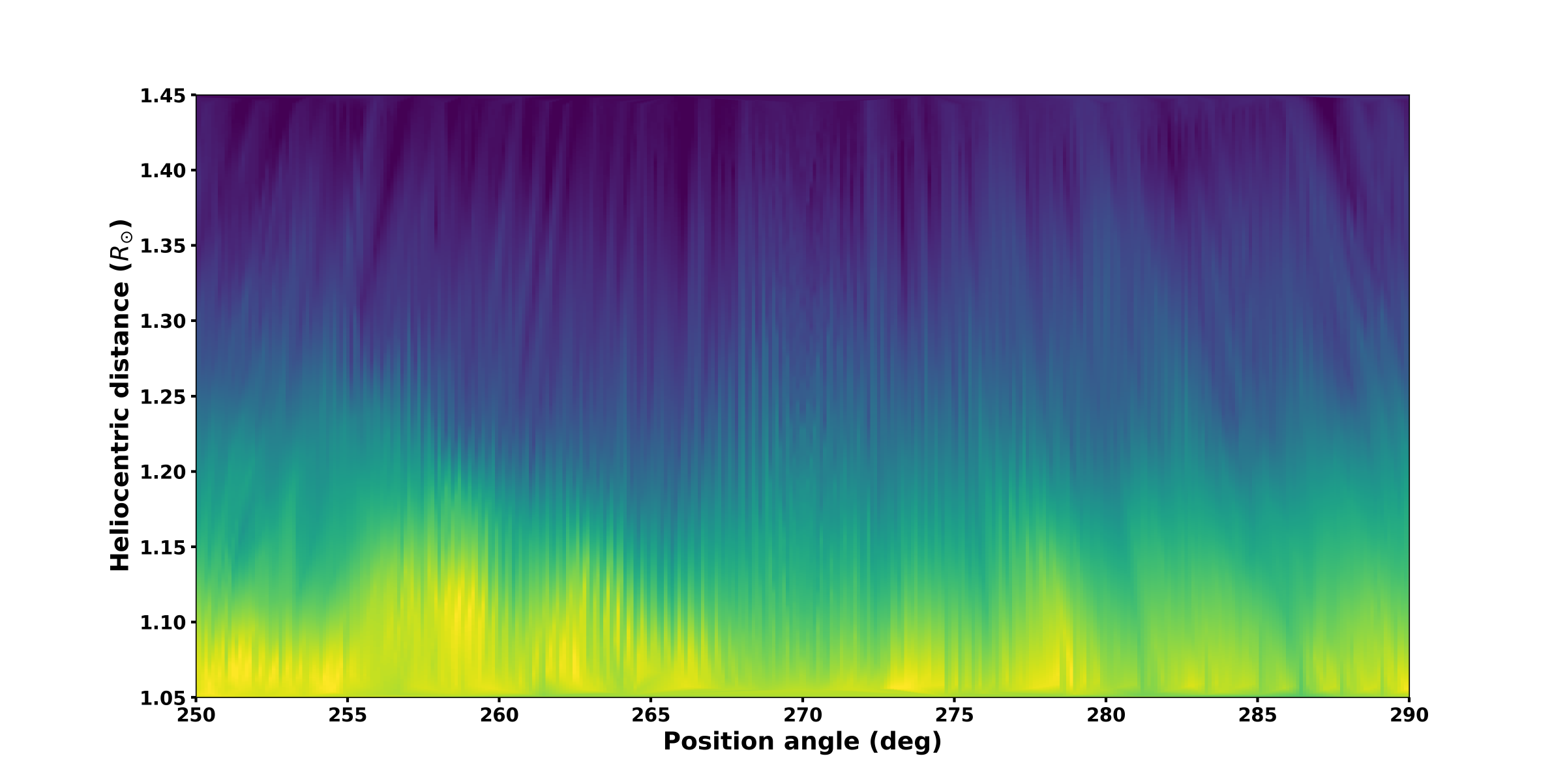}}
\caption{Same as Figure \ref{fig:figure5}, but during the interval 
${\approx}$23:09\,UT\,-\,23:56\,UT.}
\label{fig:figure6} 
\end{figure}


We also generated two intensity
images of the west limb region of Sun in 5303{\AA} emission line using the raster scan data obtained with slit 4 on 16 July 2024. These were before and after the sit \& stare observations in Figure \ref{fig:figure3}. 
The parameters related to raster scan observations mentioned in this work are listed in Table \ref{tab:table2}.
Figure \ref{fig:figure5} shows the intensity image in ($r$,${\theta}$) coordinates generated from the raster scan data obtained with slit 4 during the interval 
${\approx}$00:30\,UT\,-\,01:17\,UT. This time interval is prior to onset of the CME described above. Coronal structures extending from the edge of the occulter at 
$r$\,=\,1.05\,$\rm R_{\odot}$ 
till $r$\,${\approx}$\,1.50\,$\rm R_{\odot}$ 
could be seen in the position angle range 
${\approx}\,260^{\circ}$\,-\,$265^{\circ}$. 
Figure \ref{fig:figure6} 
shows similar map generated from the raster scan data obtained with slit 4 during the interval 
${\approx}$23:09\,UT\,-\,23:56\,UT. This time interval is after the CME described  above.
The coronal structures have changed and reduced in size in the post-CME image (Figure \ref{fig:figure6}) 
as compared to the pre-CME image ((Figure \ref{fig:figure5}).
Considering this, and the close position angle correspondence between the above mentioned coronal loop structures and the CME, we believe that the former are part of the CME that erupted. According to the SoHO/LASCO-C2 CME catalog there were other CMEs from nearly the same position angle range on 16 July 2024. But the 
H${\alpha}$ and soft X-ray flares that accompanied this particular CME were very intense as per flare classification. Furthermore, the source region of the flare was almost at the solar limb close to which the spectrograph slit 4 was positioned.

\section{Results and Discussions} \label{sec:anadis}

\begin{figure}[t!]
\centerline{\includegraphics[width=14cm]{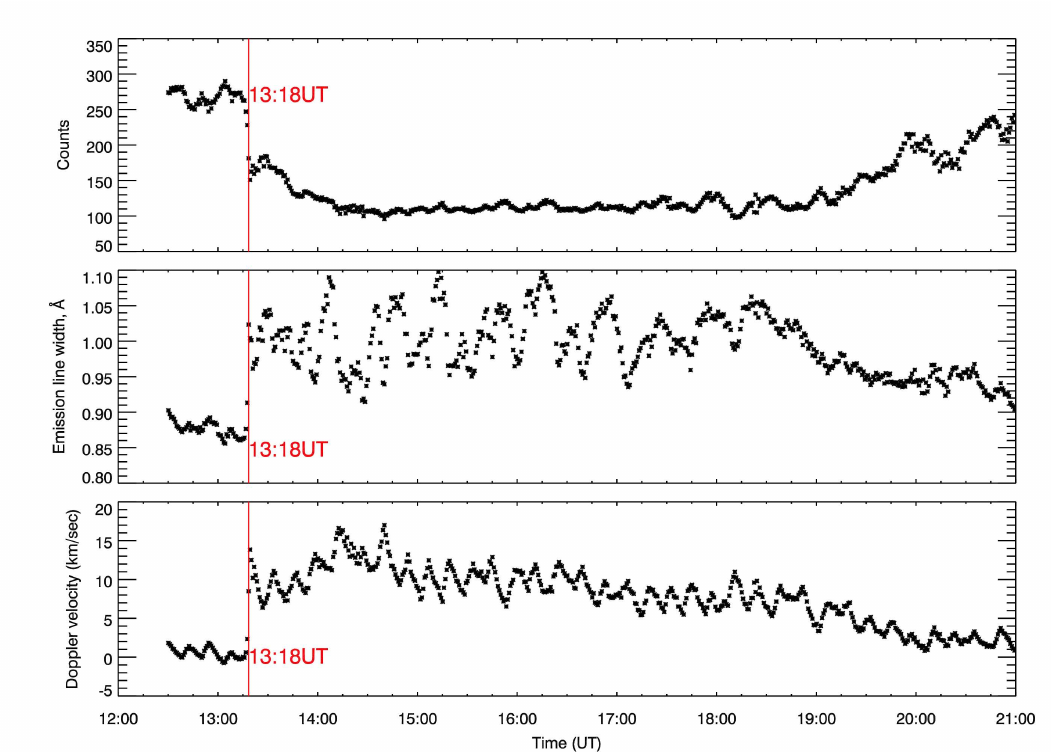}}
\caption{Upper panel: Time variation of the peak intensity of the emission line averaged over the dimming region (position angle range
${\approx}\,260^{\circ}$\,-\,$270^{\circ}$) in Figure \ref{fig:figure3}. The vertical red line indicates the onset of the dimming at 
${\approx}$\,13:18\,UT. Middle panel: Same as above, but corresponds to width of the emission line. 
Lower panel: Same as above, but corresponds to Doppler velocity estimated from the shift in the peak of the emission line from the reference.}
\label{fig:figure7}
\end{figure}

The upper panel in Figure \ref{fig:figure7} shows the time variation of 5303{\AA} emission line intensity at the dimming region 
(position angle range ${\approx}\,260^{\circ}$\,-\,$270^{\circ}$)
in Figure \ref{fig:figure3}. 
The coronal regions observed by slit 4
in the above position angle range are $r \rm {\approx}1.07\,R_{\odot}$\,-\,$\rm 1.05\,R_{\odot}$ due to the straight nature of the slit.
These distances are smaller than $r{\approx}\rm 1.2\,R_{\odot}$ upto which  
collisional excitation mechanism is believed to dominate in the line formation (see e.g. \citealp{Waldmeier1975,Singh1985,Raju1991}). Therefore, we assumed that 5303{\AA} emission line 
intensity is proportional to square of the electron density $n_{e}$.
The intensity reduction in Figure \ref{fig:figure7} is 
${\approx}$\,50\%. Therefore, the corresponding density decrease should have been 
${\approx}$\,7\%. Assuming 
$n_{e} \rm {\approx}\,1.03{\times}10^{9}cm^{-3}$ at $r$\,=\,1.05\,$\rm R_{\odot}$ in the active region corona \citep{Vrsnak2004}, we find that the above 7\% decrease in density corresponds to $\rm {\approx}\,7{\times}10^{7}\,cm^{-3}$. Note that this reduction in $n_{e}$ suggests that radiative excitation may be equally important in the line formation in the present case.
But the background $n_{e}$ is still high ($\rm {\sim}10^{9}\,cm^{-3}$) despite the decrease. The ratio of radiative to collisional excitation is expected to be minimal for such large $n_{e}$
(see e.g. \citealp{Habbal2007}). So our assumption of collisional excitation as the dominant mechanism of 5303{\AA} line formation 
in the present case remains good.
The above mentioned decrease in density agrees reasonably with similar values reported due to CME related coronal dimmings observed in X-rays and radio \citep{Sterling1997,Gopalswamy1998,Ramesh2000c,Ramesh2001b}.
The type II radio burst mentioned in Section 3 was first observed near 180\,MHz at ${\approx}$\,13:21\,UT. 
The 180\,MHz plasma level in the active region corona is expected to be  
at $r$\,${\approx}$\,1.14\,$\rm R_{\odot}$. The type II radio bursts are widely believed to be magnetohydrodynamic shocks driven by the 
leading edge of the associated CMEs (see e.g. \citealp{Ramesh2012a,Anshu2019}). Assuming the same, we estimated the possible location of the leading edge of the CME during the onset of the coronal dimming 
at 13:18\,UT. As per reports, the shock speed estimated from the type II radio burst observations on 16 July 2024 is ${\approx}$\,398\,km/s. This implies that in the interval between the onset time of the coronal 
dimming (${\approx}$13:18\,UT) and that of the type II burst (i.e. 
${\approx}$ 13:21\,UT), the CME should have propagated a distance of 
${\approx}$\,0.1\,$\rm R_{\odot}$. In other words the leading edge of the CME should have been at $r$\,${\approx}$\,1.04\,$\rm R_{\odot}$ at 13:18\,UT. There is close agreement between these numbers and that of the coronal dimming (i.e. 
$r \rm {\approx}1.05\,R_{\odot}$\,-\,$\rm 1.07\,R_{\odot}$ at 
${\approx}$13:18\,UT). The position angles of the CME (${\approx}273^{\circ}$) and that of the dimming (${\approx}265^{\circ}$) are also nearly the same. These are in support of our above argument that the dimming observed in the present case is due to CME induced depletion of coronal material. 
The duration of the dimming 
(${\approx}$6\,h) is in the range of timescales over which the corona restructures in the aftermath of a CME (see e.g. \citealp{Kathiravan2007}).

The middle panel in Figure \ref{fig:figure7} shows the variations in the 5303{\AA} emission line width ($\rm {\Delta}w$) with time, after correcting for instrumental line profile. The Doppler temperature, $\rm T_{d}$ (comprising both thermal and non-thermal components) was estimated from the emission line width using the relation, 
$\rm T_{d}\,=\,\frac{mc^{2}}{2k_{B}}\left(\frac{{\Delta}{w}}{\lambda}\right)^{2}$. Here m\,=\,$9.33{\times}10^{-26}$\,kg is the Fe XIV ion mass, and 
${\lambda}$ is the rest wavelength (5303{\AA}) of the emission line. The mean emission line widths before and after the onset of the dimming were 
${\approx}$0.87{\AA} and ${\approx}$1{\AA}, respectively. The width has increased by ${\approx}$15\% during the dimming. Substituting these values in the above relation, we find that $\rm T_{d}$ increased from 
${\approx}\,2.7{\times}10^{6}$\,K to ${\approx}\,3.6{\times}10^{6}$\,K at 13:18\,UT.
Assuming the thermal component for the Fe XIV ion, the non-thermal velocity (turbulent) was calculated using the relation  
$\rm \left[ \left(\frac{{\Delta}{w}}{\lambda}\right)^{2}\frac{c^{2}}{4ln(2)}-\left(\frac{2k_{B}T}{m}\right) \right]^{1/2}$
as ${\approx}$\,24.87\,km/s 
for $\rm {\Delta}w$\,=\,1.0{\AA}.
We assumed T\,=\,$1.8{\times}10^{6}$\,K for the above calculations. 
Enhanced turbulence because of restructuring of the coronal magnetic field and hence small-scale magnetic reconnections at the source region of the CME, after its lift-off, is a likely reason for the non-thermal line broadening in the present case.

The lower panel in Figure \ref{fig:figure7} shows the variations in Doppler velocity $\rm \left(\frac{{\Delta}{\lambda}}{\lambda}c\right)$.
The differences between the peak locations of the emission line profiles 
and the reference pixel location were computed for each spatial location along the slit, and for the entire observing duration.
$\rm {\Delta}{\lambda}$ was obtained from the above difference by  multiplying it with the spectral dispersion per pixel.
The mean Doppler velocity (averaged along the line-of-sight direction) is ${\approx}$10\,km/s during the dimming. The positive value (redshift) indicates that the 
associated plasma motions in the ambient corona are directed away from the observing direction. Note that the position angle of the CME source region (i.e. AR13738, see Section 3) is 
${\approx}264^{\circ}$. This is different from the central position angle (${\approx}273^{\circ}$) of the CME mentioned previously. 
The above difference indicates a possible deflection of the CME by the ambient magnetic field (see e.g. \citealp{Wang2020}). 
This could be a likely cause of the above mentioned redshift in the Doppler velocity though the CME originated on the Earthward side of the solar limb. The STEREO-A/EUVI observations shown in Figure \ref{fig:figure4} are consistent with the above argument related to deflection of the CME.

\section{Conclusions} \label{sec:conclu}

We have reported the first spectroscopic observations of the onset phase of a CME in the 5303{\AA} coronal emission line with the VELC payload onboard ADITYA-L1. Analysis of the peak intensity of the emission line indicate coronal dimming (${\approx}$\,50\%) in the aftermath of the CME onset. The dimming was due to depletion of coronal material because of the CME, and lasted for ${\approx}$6\,h. During the same interval, there was 
${\approx}$\,15\% enhancement in the width of the emission line. The non-thermal velocity (turbulence) during the CME 
is 
${\approx}24.87$\,km/s. Doppler shift calculations using the locations of the emission line peak during the corresponding period indicate a redshift of ${\approx}$\,10\,km/s. 
Similar near-Sun observations with VELC/ADITYA-L1 and the Upgraded Coronal Multi-channel Polarimeter (UCoMP)\footnote{https://www2.hao.ucar.edu/mlso/instruments/upgraded-coronal-multi-channel-polarimeter} can help to understand the source regions of the CMEs particularly using the data on changes in the line width and Doppler shift, the primary characteristics in the spectroscopic observations of CMEs \citep{Tian2013}. Such data can help to constrain the parameters used in the modelling of CMEs. The CME event of 16 July 2024 (associated with X1.9 class flare) described in this work was observed 
${\approx}$1\,h before it appeared in the FoV of SoHO/LASCO-C2. The planned synoptic observations with VELC/ADITYA-L1 would be very useful to constrain the onset time of the CMEs.

ADITYA-L1 is an observatory class mission which is fully funded and operated by the Indian Space Research Organization (ISRO). Data obtained with the different payloads onboard ADITYA-L1 are archived at the Indian Space Science Data Centre (ISSDC). We acknowledge the STEREO team for providing open data access. We thank the referee for his/her valuable comments which helped us to present the results more clearly.

\bibliographystyle{aasjournal}
\bibliography{VELC_ADITYAL1_Science_ApJL} 

\begin{thebibliography}{}
\expandafter\ifx\csname natexlab\endcsname\relax\def\natexlab#1{#1}\fi
\providecommand{\url}[1]{\href{#1}{#1}}
\providecommand{\dodoi}[1]{doi:~\href{http://doi.org/#1}{\nolinkurl{#1}}}
\providecommand{\doeprint}[1]{\href{http://ascl.net/#1}{\nolinkurl{http://ascl.net/#1}}}
\providecommand{\doarXiv}[1]{\href{https://arxiv.org/abs/#1}{\nolinkurl{https://arxiv.org/abs/#1}}}

\bibitem[{{Alzate} {et~al.}(2017){Alzate}, {Habbal}, {Druckm{\" u}ller},
  {Emmanouilidis}, \& {Morgan}}]{Alzate2017}
{Alzate}, N., {Habbal}, S., {Druckm{\" u}ller}, M., {Emmanouilidis}, C., \&
  {Morgan}, H. 2017, \apj, 848, 84, \dodoi{10.3847/1538-4357/aa8cd2}

\bibitem[{{Bagala} {et~al.}(2001){Bagala}, {Stenborg}, {Schwenn}, \&
  {Haerendel}}]{Bagala2001}
{Bagala}, L.~G., {Stenborg}, G., {Schwenn}, R., \& {Haerendel}, G. 2001, \jgr,
  106, 25239, \dodoi{10.1029/2000JA004017}

\bibitem[{{Boe} {et~al.}(2020){Boe}, {Habbal}, {Druckm{\" u}ller}, {Ding},
  {Hod{\' e}rov}, \& {{\v S}tarha}}]{Boe2020}
{Boe}, B., {Habbal}, S., {Druckm{\" u}ller}, M., {et~al.} 2020, \apj, 888, 100,
  \dodoi{10.3847/1538-4357/ab5e34}

\bibitem[{{Brueckner} {et~al.}(1995){Brueckner}, {Howard}, {Koomen},
  {Korendyke}, {Michels}, {Moses}, {Socker}, {Dere}, {Lamy}, {Llebaria},
  {Bout}, {Schwenn}, {Simnett}, {Bedford}, \& {Eyles}}]{Brueckner1995}
{Brueckner}, G.~E., {Howard}, R.~A., {Koomen}, M.~J., {et~al.} 1995, \solphys,
  162, 357, \dodoi{10.1007/BF00733434}

\bibitem[{{Bruzek} \& {Demastus}(1970)}]{Bruzek1970}
{Bruzek}, A., \& {Demastus}, H.~L. 1970, \solphys, 12, 447,
  \dodoi{10.1007/BF00148027}

\bibitem[{{Caspi} {et~al.}(2020){Caspi}, {Seaton}, {Tsang}, {DeForest},
  {Bryans}, {DeLuca}, \& {Tomczyk}}]{Caspi2020}
{Caspi}, A., {Seaton}, D.~B., {Tsang}, C. C.~C., {et~al.} 2020, \apj, 702, 131,
  \dodoi{10.3847/1538-4357/ab89a8}

\bibitem[{{Demastus} {et~al.}(1973){Demastus}, {Wagner}, \&
  {Robinson}}]{Demastus1973}
{Demastus}, H.~L., {Wagner}, W.~J., \& {Robinson}, R.~D. 1973, \solphys, 31,
  449, \dodoi{10.1007/BF00152820}

\bibitem[{{Dere} {et~al.}(1997){Dere}, {Brueckner}, {Howard}, {Koomen},
  {Korendyke}, \& {Kreplin}}]{Dere1997}
{Dere}, K.~P., {Brueckner}, G.~E., {Howard}, R.~A., {et~al.} 1997, \solphys,
  175, 601, \dodoi{10.1023/A:1004907307376}

\bibitem[{{Evans}(1957)}]{Evans1957}
{Evans}, J.~W. 1957, Publ. Astron. Soc. Pacific, 69, 421,
  \dodoi{10.1086/127116}

\bibitem[{{Gopalswamy} \& {Hanaoka}(1998)}]{Gopalswamy1998}
{Gopalswamy}, N., \& {Hanaoka}, Y. 1998, \apj, 498, L179,
  \dodoi{10.1086/311330}

\bibitem[{{Guhathakurta} {et~al.}(1993){Guhathakurta}, {Fisher}, \&
  {Altrock}}]{Guhathakurta1993}
{Guhathakurta}, M., {Fisher}, R.~R., \& {Altrock}, R.~C. 1993, \apj, 414, L145,
  \dodoi{10.1086/187017}

\bibitem[{{Habbal} {et~al.}(2007){Habbal}, {Morgan}, {Johnson}, {Arndt}, {Daw},
  {Jaeggli}, {Kuhn}, \& {Mickey}}]{Habbal2007}
{Habbal}, S.~R., {Morgan}, H., {Johnson}, J., {et~al.} 2007, \apj, 663, 598,
  \dodoi{10.1086/518403}

\bibitem[{{Hori} {et~al.}(2005){Hori}, {Ichimoto}, {Sakurai}, {Sano}, \&
  {Nishino}}]{Hori2005}
{Hori}, K., {Ichimoto}, K., {Sakurai}, T., {Sano}, I., \& {Nishino}, Y. 2005,
  \apj, 618, 1001, \dodoi{10.1086/426013}

\bibitem[{{Howard} {et~al.}(2008){Howard}, {Moses}, {Vourlidas}, {Newmark},
  {Socker}, {Plunkett}, {Korendyke}, {Cook}, {Hurley}, {Davila}, {Thompson},
  {St Cyr}, {Mentzell}, {Mehalick}, {Lemen}, {Wuelser}, {Duncan}, {Tarbell},
  {Wolfson}, {Moore}, {Harrison}, {Waltham}, {Lang}, {Davis}, {Eyles},
  {Mapson-Menard}, {Simnett}, {Halain}, {Defise}, {Mazy}, {Rochus}, {Mercier},
  {Ravet}, {Delmotte}, {Auchere}, {Delaboudiniere}, {Bothmer}, {Deutsch},
  {Wang}, {Rich}, {Cooper}, {Stephens}, {Maahs}, {Baugh}, {McMullin}, \&
  {Carter}}]{Howard2008}
{Howard}, R.~A., {Moses}, J.~D., {Vourlidas}, A., {et~al.} 2008, \ssr, 136, 67,
  \dodoi{10.1007/s11214-008-9341-4}

\bibitem[{{Ichimoto} {et~al.}(1995{\natexlab{a}}){Ichimoto}, {Hara}, {Takeda},
  {Kumagai}, {Sakurai}, {Shimizu}, \& {Hudson}}]{Ichimoto1995b}
{Ichimoto}, K., {Hara}, H., {Takeda}, A., {et~al.} 1995{\natexlab{a}}, \apj,
  445, 978, \dodoi{10.1086/175756}

\bibitem[{{Ichimoto} {et~al.}(1995{\natexlab{b}}){Ichimoto}, {Ohtani},
  {Ishigaki}, {Maemura}, \& {Noguchi}}]{Ichimoto1995a}
{Ichimoto}, K., {Ohtani}, H., {Ishigaki}, T., {Maemura}, H., \& {Noguchi}, M.
  1995{\natexlab{b}}, Publ. Astron. Soc. Japan, 47, 383

\bibitem[{{Inhester} \& {Schwenn}(1997)}]{Inhester1997}
{Inhester}, B., \& {Schwenn}, R. 1997, in Correlated Phenomena at the Sun, in
  the Heliosphere and in Geospace, ed. A.~{Wilson}, European Space Agency, ESA
  SP-415, 47

\bibitem[{{Kathiravan} {et~al.}(2007){Kathiravan}, {Ramesh}, \&
  {Nataraj}}]{Kathiravan2007}
{Kathiravan}, C., {Ramesh}, R., \& {Nataraj}, H.~S. 2007, \apj, 656, L37,
  \dodoi{10.1086/512013}

\bibitem[{{Koutchmy} {et~al.}(1983){Koutchmy}, {Zhugzhda}, \&
  {Locans}}]{Koutchmy1983}
{Koutchmy}, S., {Zhugzhda}, I.~D., \& {Locans}, V. 1983, \aap, 120, 185

\bibitem[{{Kumari} {et~al.}(2019){Kumari}, {Ramesh}, {Kathiravan}, {Wang}, \&
  {Gopalswamy}}]{Anshu2019}
{Kumari}, A., {Ramesh}, R., {Kathiravan}, C., {Wang}, T.~J., \& {Gopalswamy},
  N. 2019, \apj, 881, 24, \dodoi{10.3847/1538-4357/ab2adf}

\bibitem[{{Landi} {et~al.}(2016){Landi}, {Habbal}, \& {Tomczyk}}]{Landi2016}
{Landi}, E., {Habbal}, S.~R., \& {Tomczyk}, S. 2016, J. Geophys. Res.: Space
  Physics, 121, 8237, \dodoi{10.1002/2016JA022598}

\bibitem[{{Mierla} {et~al.}(2008){Mierla}, {Schwenn}, {Teriaca}, {Stenborg}, \&
  {Podlipnik}}]{Mierla2008}
{Mierla}, M., {Schwenn}, R., {Teriaca}, L., {Stenborg}, G., \& {Podlipnik}, B.
  2008, \aap, 480, 509, \dodoi{10.1051/0004-6361:20078329}

\bibitem[{{Minarovjech} {et~al.}(2003){Minarovjech}, {Ru\v{s}in},
  {Rybansk\'{y}}, {Sakurai}, \& {Ichimoto}}]{Minarovjech2003}
{Minarovjech}, M., {Ru\v{s}in}, V., {Rybansk\'{y}}, M., {Sakurai}, T., \&
  {Ichimoto}, K. 2003, \solphys, 213, 269, \dodoi{10.1023/A:1023938732756}

\bibitem[{{Mishra} {et~al.}(2024){Mishra}, {Sasikumar Raja}, {Sanal Krishnan},
  {Suresh Venkata}, {Bhavana Hegde}, \& {Utkarsha}}]{Mishra2024}
{Mishra}, S., {Sasikumar Raja}, K., {Sanal Krishnan}, V.~U., {et~al.} 2024,
  Exp. Astron., 57, 7, \dodoi{10.1007/s10686-024-09922-2}

\bibitem[{{Muro} {et~al.}(2023){Muro}, {Gunn}, {Fearn}, {Fearn}, \&
  {Morgan}}]{Muro2023}
{Muro}, G.~D., {Gunn}, M., {Fearn}, S., {Fearn}, T., \& {Morgan}, H. 2023,
  \solphys, 298, 75, \dodoi{10.1007/s11207-023-02162-1}

\bibitem[{{Muthu Priyal} {et~al.}(2024){Muthu Priyal}, {Singh}, {Prasad},
  {Sumana}, {Varun Kumar}, \& {Mishra}}]{Muthupriyal2024}
{Muthu Priyal}, V., {Singh}, J., {Prasad}, B.~R., {et~al.} 2024, Adv. Space
  Res., 74, 547, \dodoi{10.1016/j.asr.2024.03.058}

\bibitem[{{Orrall} \& {Smith}(1961)}]{Orrall1961}
{Orrall}, F.~Q., \& {Smith}, H.~J. 1961, \aj, 66, 293, \dodoi{10.1086/108425}

\bibitem[{{Pasachoff} {et~al.}(2002){Pasachoff}, {Babcock}, {Russell}, \&
  {Seaton}}]{Pasachoff2002}
{Pasachoff}, J.~M., {Babcock}, B.~A., {Russell}, K.~D., \& {Seaton}, D.~B.
  2002, \solphys, 207, 241, \dodoi{10.1023/A:1016297800478}

\bibitem[{{Pasachoff} {et~al.}(2009){Pasachoff}, {Ru{\v s}in}, {Druckm{\"
  u}ller}, {Aniol}, {Saniga}, \& {Minarovjech}}]{Pasachoff2009}
{Pasachoff}, J.~M., {Ru{\v s}in}, V., {Druckm{\" u}ller}, M., {et~al.} 2009,
  \apj, 702, 1297, \dodoi{10.1088/0004-637X/702/2/1297}

\bibitem[{{Plunkett} {et~al.}(1997){Plunkett}, {Brueckner}, {Dere}, {Howard},
  {Koomen}, \& {Korendyke}}]{Plunkett1997}
{Plunkett}, S.~P., {Brueckner}, G.~E., {Dere}, K.~P., {et~al.} 1997, \solphys,
  175, 699, \dodoi{10.1023/A:1004981125702}

\bibitem[{{Prasad} {et~al.}(2017){Prasad}, {Banerjee}, {Singh}, {Nagabhushana},
  {Amit Kumar}, \& {Kamath}}]{Prasad2017}
{Prasad}, B.~R., {Banerjee}, D., {Singh}, J., {et~al.} 2017, Curr. Sci., 113,
  613.
\newblock \url{http://www.jstor.org/stable/26293898}

\bibitem[{{Prasad} {et~al.}(2023){Prasad}, {Suresh Venkata}, {Natarajan},
  {Pawan Kumar}, {Kamath}, \& {Mishra}}]{Prasad2023}
{Prasad}, B.~R., {Suresh Venkata}, N., {Natarajan}, V., {et~al.} 2023, J.
  Astron. Teles. Inst. Sys., 9, 044001

\bibitem[{{Raj Kumar} {et~al.}(2018){Raj Kumar}, {Prasad}, {Singh}, \& {Suresh
  Venkata}}]{Rajkumar2018}
{Raj Kumar}, N., {Prasad}, B.~R., {Singh}, J., \& {Suresh Venkata}, N. 2018,
  Exp. Astron., 45, 219, \dodoi{10.1007/s10686-017-9569-7}

\bibitem[{{Raju} {et~al.}(1991){Raju}, {Desai}, , {Chandrasekhar}, \&
  {Ashok}}]{Raju1991}
{Raju}, K.~P., {Desai}, J.~N., , {Chandrasekhar}, T., \& {Ashok}, N.~M. 1991,
  J. Astron. Astrophys., 12, 311, \dodoi{10.1007/BF02702319}

\bibitem[{{Ramesh} {et~al.}(2012){Ramesh}, {Anna Lakshmi}, {Kathiravan},
  {Gopalswamy}, \& {Umapathy}}]{Ramesh2012a}
{Ramesh}, R., {Anna Lakshmi}, M., {Kathiravan}, C., {Gopalswamy}, N., \&
  {Umapathy}, S. 2012, \apj, 752, 107, \dodoi{10.1088/0004-637X/752/2/107}

\bibitem[{{Ramesh} \& {Ebenezer}(2001)}]{Ramesh2001b}
{Ramesh}, R., \& {Ebenezer}, E. 2001, \apjl, 558, L141, \dodoi{10.1086/323498}

\bibitem[{{Ramesh} \& {Sastry}(2000)}]{Ramesh2000c}
{Ramesh}, R., \& {Sastry}, C.~V. 2000, \aap, 358, 749

\bibitem[{{Ru\v{s}in} {et~al.}(1994){Ru\v{s}in}, {Rybansk\'{y}}, {Minarovjech},
  \& {Pinter}}]{Rusin1995}
{Ru\v{s}in}, V., {Rybansk\'{y}}, M., {Minarovjech}, M., \& {Pinter}, T. 1994,
  in Infrared solar physics, ed. D.~M. {Rabin}, J.~T. {Jefferies}, \&
  C.~{Lindsey}, IAU Symp. 154, 211

\bibitem[{{Sakurai}(1998)}]{Sakurai1998}
{Sakurai}, T. 1998, in Synoptic Solar Physics, ed. K.~S. {Balasubramaniam},
  J.~{Harvey}, \& J.~{Rabin}, ASP Conf. Ser., Vol.140, 483

\bibitem[{{Schwenn} {et~al.}(1997){Schwenn}, {Inhester}, {Plunkett}, {Epple},
  {Podlipnik}, \& {Bedford}}]{Schwenn1997}
{Schwenn}, R., {Inhester}, B., {Plunkett}, S.~P., {et~al.} 1997, \solphys, 175,
  667, \dodoi{10.1023/A:1004948913883}

\bibitem[{{Singh}(1985)}]{Singh1985}
{Singh}, J. 1985, \solphys, 95, 253, \dodoi{10.1007/BF00152403}

\bibitem[{{Singh} {et~al.}(2011{\natexlab{a}}){Singh}, {Hasan}, {Gupta},
  {Nagaraju}, \& {Banerjee}}]{Singh2011}
{Singh}, J., {Hasan}, S.~S., {Gupta}, G.~R., {Nagaraju}, K., \& {Banerjee}, D.
  2011{\natexlab{a}}, \solphys, 270, 213, \dodoi{10.1007/s11207-011-9732-7}

\bibitem[{{Singh} {et~al.}(1999){Singh}, {Ichimoto}, {Imai}, {Sakurai}, \&
  {Takeda}}]{Singh1999}
{Singh}, J., {Ichimoto}, K., {Imai}, H., {Sakurai}, T., \& {Takeda}, A. 1999,
  Publ. Astron. Soc. Japan, 51, 269, \dodoi{10.1093/pasj/51.2.269}

\bibitem[{{Singh} {et~al.}(2019){Singh}, {Prasad}, {Suresh Venkata}, \& {Amit
  Kumar}}]{Singh2019}
{Singh}, J., {Prasad}, B.~R., {Suresh Venkata}, N., \& {Amit Kumar}. 2019, Adv.
  Space Res., 64, 1455, \dodoi{10.1016/j.asr.2019.07.007}

\bibitem[{{Singh} {et~al.}(2011{\natexlab{b}}){Singh}, {Prasad},
  {Venkatakrishnan}, {Sankarasubramanian}, {Banerjee}, \& {Raja
  Bayanna}}]{Singh2011a}
{Singh}, J., {Prasad}, B.~R., {Venkatakrishnan}, P., {et~al.}
  2011{\natexlab{b}}, Curr. Sci., 100, 167

\bibitem[{{Singh} {et~al.}(2024){Singh}, {Ramesh}, {Prasad}, {Muthu Priyal},
  {Sasikumar Raja}, \& {Suresh Venkata}}]{Singh2024}
{Singh}, J., {Ramesh}, R., {Prasad}, B.~R., {et~al.} 2024, \solphys~(submitted)

\bibitem[{{Singh} {et~al.}(2004){Singh}, {Sakurai}, {Ichimoto}, \&
  {Watanabe}}]{Singh2004}
{Singh}, J., {Sakurai}, T., {Ichimoto}, K., \& {Watanabe}, T. 2004, \apj, 617,
  L81, \dodoi{10.1086/427078}

\bibitem[{{Sterling} \& {Hudson}(1997)}]{Sterling1997}
{Sterling}, A.~C., \& {Hudson}, H.~S. 1997, \apj, 491, L55,
  \dodoi{10.1086/311043}

\bibitem[{{Suresh Venkata} \& {Prasad}(2021)}]{Venkata2021}
{Suresh Venkata}, N., \& {Prasad}, B.~R. 2021, Opt. Engg., 60, 074103,
  \dodoi{10.1117/1.OE.60.7.074103}

\bibitem[{{Suresh Venkata} {et~al.}(2017){Suresh Venkata}, {Prasad}, {Raj
  Kumar}, \& {Singh}}]{Venkata2017}
{Suresh Venkata}, N., {Prasad}, B.~R., {Raj Kumar}, N., \& {Singh}, J. 2017, J.
  Astron. Teles. Inst. Sys., 3, 014002, \dodoi{10.1117/1.JATIS.3.1.014002}

\bibitem[{{Suzuki} {et~al.}(2006){Suzuki}, {Sakurai}, \&
  {Ichimoto}}]{Suzuki2006}
{Suzuki}, I., {Sakurai}, T., \& {Ichimoto}, K. 2006, Publ. Astron. Soc. Japan,
  58, 165, \dodoi{10.1093/pasj/58.1.165}

\bibitem[{{Sykora}(1992)}]{Sykora1992}
{Sykora}, J. 1992, \solphys, 140, 379, \dodoi{10.1007/BF00146319}

\bibitem[{{Tian} {et~al.}(2013){Tian}, {Tomczyk}, {McIntosh}, {Bethge}, {de
  Toma}, \& {Gibson}}]{Tian2013}
{Tian}, H., {Tomczyk}, S., {McIntosh}, S.~W., {et~al.} 2013, \solphys, 288,
  637, \dodoi{10.1007/s11207-013-0317-5}

\bibitem[{{Tsubaki}(1975)}]{Tsubaki1975}
{Tsubaki}, T. 1975, \solphys, 43, 147, \dodoi{10.1007/BF00155151}

\bibitem[{{Voulgaris} {et~al.}(2012){Voulgaris}, {Gaintatzis}, {Seiradakis},
  {Pasachoff}, \& {Economou}}]{Voulgaris2012}
{Voulgaris}, A.~G., {Gaintatzis}, P.~S., {Seiradakis}, J.~S., {Pasachoff},
  J.~M., \& {Economou}, T.~E. 2012, \solphys, 278, 187,
  \dodoi{10.1007/s11207-012-9929-4}

\bibitem[{{Vr{\v s}nak} {et~al.}(2004){Vr{\v s}nak}, {Magdaleni{\'c}}, \&
  {Zlobec}}]{Vrsnak2004}
{Vr{\v s}nak}, B., {Magdaleni{\'c}}, J., \& {Zlobec}, P. 2004, \aap, 413, 753,
  \dodoi{10.1051/0004-6361:20034060}

\bibitem[{{Waldmeier}(1975)}]{Waldmeier1975}
{Waldmeier}, M. 1975, \solphys, 45, 147, \dodoi{10.1007/BF00152227}

\bibitem[{{Wang} {et~al.}(2020){Wang}, {Hoeksema}, \& {Liu}}]{Wang2020}
{Wang}, J., {Hoeksema}, J.~T., \& {Liu}, S. 2020, J. Geophys. Res.: Space
  Physics, 125, e27530, \dodoi{10.1029/2019JA027530}

\end{thebibliography}

\end{document}